# Geometrodynamics of polarized light:
# Berry phase and spin Hall effect in a gradient-index medium

Konstantin Y. Bliokh[1,2,3*]

[1]*Nonlinear Physics Centre, Research School of Physics and Engineering,
The Australian National University, Canberra ACT 0200, Australia*
[2]*Micro- and Nanooptics Laboratory, Faculty of Mechanical Engineering,
Technion–Israel Institute of Technology, Haifa 32000, Israel*
[3]*Institute of Radio Astronomy, 4 Krasnoznamyonnaya Str., Kharkov 61002, Ukraine*

**Abstract.** We review the geometrical-optics evolution of an electromagnetic wave propagating along a curved ray trajectory in a gradient-index dielectric medium. A Coriolis-type term appears in Maxwell equations under transition to the rotating coordinate system accompanying the ray. This term describes the spin-orbit coupling of light which consists of (i) the Berry phase responsible for a trajectory-dependent polarization variations and (ii) the spin Hall effect representing polarization-dependent trajectory perturbations. These mutual phenomena are described within universal geometrical structures underlying the problem and are explained by the dynamics of the intrinsic angular momentum carried by the wave. Such close geometro-dynamical interrelations illuminate a dual physical nature of the phenomena.

## 1. Introduction

Spin dynamics of quantum particles, which is induced by weak spin-orbit interactions, is currently attracting growing attention. This is owing to both fundamental theoretical interest and potential applications in atomic, condensed-matter, and optical systems at nano-scales. The spin-orbit interaction is an inherent topological feature of relativistic wave equations such as, e.g., Dirac or Maxwell equations, and spin and orbital degrees of freedom of the particle become coupled as the particle evolves in an external potential. In the short-wavelength (semiclassical) approximation, the spin-orbit coupling can be described within the Berry-phase formalism, which has a universal geometrical character and can be applied to a diversity of quantum and classical wave systems [1–3].

Geometrical optics embraces all features of the semiclassical evolution of relativistic quantum particles with spin and, thus, offers a unique opportunity for investigations of the fundamental quantum effects and applications within the classical field. In so doing, the evolution of polarized light in a gradient-index medium mimics evolution of a massless spin-1 particle (photon) in an external scalar field. The Berry phase provides a natural *geometrical* formalism that describes variations of the light polarization along the propagation trajectory (see reviews [4–7] and original papers [8–20]).[1] However, the Berry phase also manifests itself *dynamically*. In particular, it induces an additional term in the equations of motion (ray equations) which describes a polarization-dependent shift of the wave trajectory. This effect is known as the *spin Hall effect of light* or optical Magnus effect [24–35] which is quite similar to the intrinsic anomalous and spin-Hall effects in quantum systems [29,36–40] (for reviews, see [41–43]). Together, the Berry phase and spin Hall effect describe the *spin-orbital coupling* of electromagnetic waves, i.e., a mutual influence of the wave trajectory and polarization. In a more general context, the Berry phase and the accompanying geometric reaction forces arise from a coupling between the fast and slow degrees of freedom [44] (for other examples of the topological transport phenomena, see [45–58]).

---

[*]E-mail: k.bliokh@gmail.com
[1] In this review, we consider only the spin-redirection Rytov–Vladimirskii–Berry phase in inhomogeneous media, which is associated with variations in the direction of propagation of light. Another geometric phase, the Pancharatnam–Berry phase [21–23], may appear in anisotropic media; it is caused by changes in the polarization state of light (see reviews [4–7]).

In this review, we consider the Berry-phase phenomena upon the geometrical-optics evolution of light in a gradient-index medium. We aim to perform a thorough analysis of this problem but will not cover its important generalizations to step-index interfaces, anisotropic media, and higher-order states of light carrying orbital angular momentum. Paying proper attention to the inherent geometrical features, we particularly focus on the dynamical aspects of the problem. It will be shown that the Berry phase and spin Hall effect are closely related to the dynamics of the *spin angular momentum* carried by the wave and can be described in terms of the Coriolis effect or the angular Doppler effect. To reveal interrelations between various formalisms, we attempt to give different interpretations of the same results.

The paper is organized as follows. Section 2 provides a brief introduction to the scalar geometrical optics of gradient-index media. The main Section 3 describes various aspects of the polarization evolution of electromagnetic waves, including the Berry phase and spin Hall effect. Section 4 analyses a typical example of these phenomena – a helical ray trajectory in a cylindrical medium. In the concluding Section 5 we discuss basic physical interpretations of the Berry phase and spin Hall effect and mention the most important generalizations. Specific coordinate frames, which are used for analysis of the polarization evolution along the ray trajectory, are discussed in the Appendix.

## 2. Geometrical optics of scalar waves

Geometrical optics of gradient-index media is an asymptotic short-wavelength approximation for the solution of the wave equation, which describes propagation of a paraxial wave packet or beam with nearly plane phase front [59]. In this approximation, the wave acquires features of a classical point particle obeying Hamiltonian or Lagrangian dynamics. For a scalar monochromatic wave of the frequency $\omega$, the wave equation is reduced to the Helmholz equation:

$$\left( \hbar_0^2 \nabla^2 + n^2 \right) \psi = 0 , \qquad (2.1)$$

where $\hbar_0 = \lambda_0 / 2\pi$, $\lambda_0 = c/\omega$ is the wavelength in vacuum, $n = n(\mathbf{r})$ is the space-variant refractive index of the medium, and $\psi$ is the wave field. We assume that inhomogeneity of the medium is smooth, i.e., its characteristic scale, $L$, is large as compared with the wavelength:

$$\mu = \frac{\hbar}{L} \ll 1 , \qquad (2.2)$$

where $\mu$ is the small parameter of geometrical optics, $\hbar = \hbar_0 / n$ is the wavelength in the medium, and $L \sim n/|\nabla n|$. Asymptotics of Eq. (2.1) with respect to the small parameter (2.2) is similar to the semiclassical asymptotics $\hbar \to 0$ in quantum wave equations. In the zero-order approximation, i.e., in the 'classical' limit $\hbar_0 \to 0$, propagation of wave beams is described by canonical equations with Hamiltonian [59]

$$\mathcal{H}_0(\mathbf{p}, \mathbf{r}) = \alpha \left[ p^2 - n^2(\mathbf{r}) \right] = 0 . \qquad (2.3)$$

Here $\mathbf{p} = \hbar_0 \mathbf{k}$ is the dimensionless momentum of the wave, $\mathbf{k}$ is the central wave vector in the beam, and $\alpha$ is a non-zero factor which determines the parameterization of the ray trajectory. Hamiltonian (2.3) follows from Eq. (2.1) via 'quantum-to-classical' substitution $-i\hbar_0 \nabla \to \mathbf{p}$.

Condition $p^2 = n^2$ is the local dispersion relation; it plays the role of a constraint. It is convenient to choose $\alpha = (p+n)^{-1}$ which yields

$$\mathcal{H}_0(\mathbf{p},\mathbf{r}) = p - n(\mathbf{r}) = 0. \tag{2.4}$$

The equations of motion – the ray equations – with Hamiltonian (2.4) read [59]

$$\dot{\mathbf{p}} = -\frac{\partial \mathcal{H}_0}{\partial \mathbf{r}} = \nabla n, \quad \dot{\mathbf{r}} = \frac{\partial \mathcal{H}_0}{\partial \mathbf{p}} = \frac{\mathbf{p}}{p}. \tag{2.5}$$

Hereafter the overdot stands for the derivative with respect to the ray parameter, $l$, which is the ray length since $|\dot{\mathbf{r}}| = 1$. Equations (2.5) describe *rays*, i.e., characteristics of the initial wave equation (2.1) and trajectories of evolution of the center of gravity of a paraxial wave packet or beam. The rays are curvilinear because of the wave refraction in the medium, and the gradient $\nabla n$ plays the role of an 'external force' affecting the motion of the particle. Alternatively, canonical equations (2.5) can be represented as the Euler–Lagrange equations with the Lagrangian

$$\mathcal{L}_0(\mathbf{p},\dot{\mathbf{p}},\mathbf{r},\dot{\mathbf{r}}) = -\mathcal{H}_0(\mathbf{p},\mathbf{r}) + \mathbf{p}\dot{\mathbf{r}} \tag{2.6}$$

varying independently with respect to $\mathbf{p}$ and $\mathbf{r}$.

The geometrical optics describes not only 'classical', particle-like features of the solution but also basic wave characteristics. In particular, the central wave *phase* $\Phi_0$ ($\psi \propto e^{i\Phi_0}$) in this approximation is given by the WKB integral along the ray:

$$\Phi_0 = \bar{\lambda}_0^{-1} \int \mathcal{L}\, dl = \bar{\lambda}_0^{-1} \int_L \mathbf{p}\, d\mathbf{r} = \int_L \mathbf{k}\, d\mathbf{r}, \tag{2.7}$$

where $L$ is the ray trajectory, Eqs. (2.5), in the real space, so that integral (2.7) equals $\Phi_0 = \int k\, dl = \bar{\lambda}_0^{-1} \int n\, dl$.

# 3. Geometrical optics of vector waves: spin-orbit geometrodynamics

Evolution of vector waves, e.g., electromagnetic or elastic ones, requires more sophisticated approaches, since it involves an additional, internal degree of freedom – *polarization* (spin).

## 3.1. *Maxwell equations and polarization orthogonality*

Maxwell equations for monochromatic electric field in an inhomogeneous isotropic dielectric medium can be written as

$$-\bar{\lambda}_0^2 \nabla \times (\nabla \times \mathbf{E}) + n^2 \mathbf{E} = 0,$$

or,

$$\left(\bar{\lambda}_0^2 \nabla^2 + n^2\right)\mathbf{E} - \bar{\lambda}_0^2 \nabla(\nabla \mathbf{E}) = 0, \tag{3.1}$$

where $n^2 = \varepsilon$ is the dielectric constant of the medium, which is assumed to be real. Equation (3.1) resembles the Helmholz equation (2.1), except for the last term, which involves the wave polarization, and mixes internal and external degrees of freedom of the wave. This polarization term is small in a gradient-index medium (of the order of $\mu$) and can be treated perturbatively within the geometrical optics approach. In the *zero*-order approximation in $\mu$, Eq. (3.1) results in the scalar Hamiltonian and the ray equations (2.3)–(2.6). However, polarization is indeterminate in this approximation. To describe the polarization evolution of electromagnetic waves, one has to solve the problem in the *first*-order approximation in $\mu$ [59], which accounts for the spin-orbit coupling of light. The $\mu$-order corrections describes dynamics of the wave polarization along the zero-order trajectory (2.5) as well as the polarization-dependent perturbations of the ray trajectory.

The polarization term in Eq. (3.1) ensures the Maxwell equation
$$\nabla(n^2 \mathbf{E}) = 0, \qquad (3.2)$$
i.e., $\text{div}\, \mathbf{D} = 0$, which indicates that in a smoothly inhomogeneous medium the electric field of a paraxial wave remains nearly *transverse* with respect to the current momentum $\mathbf{p}$:
$$\mathbf{E} = \mathbf{E}_\perp + E_\parallel \mathbf{t}, \quad \mathbf{E}_\perp \perp \mathbf{t}, \quad |E_\parallel| \ll |E_\perp|. \qquad (3.3)$$
Here $\mathbf{t} = \mathbf{p}/p$ is the unit vector tangent to the zero-order ray trajectory (2.5), $E_\parallel = \mathbf{E}\mathbf{t}$ is the longitudinal component of the field, and $\mathbf{E}_\perp$ is the projection of the electric field onto the plane orthogonal to $\mathbf{t}$: $\mathbf{E}_\perp = \mathbf{E} - E_\parallel \mathbf{t}$. The wave polarization is essentially determined by the transverse components $\mathbf{E}_\perp$. Hence, the dimension of the problem can be reduced to 2 by *projecting* Maxwell equation (3.1) onto the plane orthogonal to $\mathbf{t}$, which eliminates the longitudinal field component $E_\parallel$ from the problem. Below we perform this procedure and describe the evolution of $\mathbf{E}_\perp$.

## 3.2. Coriolis term in the co-moving frame

In order to describe the evolution of the transverse field (3.3), one has to introduce a coordinate frame with basis vectors $(\mathbf{v}, \mathbf{w}, \mathbf{t})$ attached to the local direction of momentum, $\mathbf{t}$, see Fig. 1. Vectors $(\mathbf{v}, \mathbf{w})$ provide a natural basis for linear polarizations: $\mathbf{E}_\perp = E_v \mathbf{v} + E_w \mathbf{w}$. However, the coordinate frame $(\mathbf{v}, \mathbf{w}, \mathbf{t})$ is *non-inertial* in the generic case. Indeed, it experiences rotation as $\mathbf{t}$ varies along the ray trajectory (2.5) in inhomogeneous medium. Such rotation with respect to a laboratory (motionless) coordinate frame is described by a precession of the triad $(\mathbf{v}, \mathbf{w}, \mathbf{t})$ with some 'angular velocity' $\boldsymbol{\Omega}$ defined with respect to the ray length $l$:
$$\dot{\mathbf{v}} = \boldsymbol{\Omega} \times \mathbf{v}, \quad \dot{\mathbf{w}} = \boldsymbol{\Omega} \times \mathbf{w}, \quad \dot{\mathbf{t}} = \boldsymbol{\Omega} \times \mathbf{t}. \qquad (3.4)$$
The angular velocity can be written as
$$\boldsymbol{\Omega} = (\dot{\mathbf{v}}\mathbf{w})\mathbf{t} + (\dot{\mathbf{w}}\mathbf{t})\mathbf{v} + (\dot{\mathbf{t}}\mathbf{v})\mathbf{w} = \Omega_\parallel \mathbf{t} + \mathbf{t} \times \dot{\mathbf{t}}, \qquad (3.5)$$
where $\Omega_\parallel = \boldsymbol{\Omega}\mathbf{t} = \dot{\mathbf{v}}\mathbf{w}$ is the longitudinal component of $\boldsymbol{\Omega}$.

When performing a transition to the non-inertial coordinate frame accompanying the ray, effective inertia terms appear in Maxwell equations (3.1). Similarly to classical mechanics, they can be derived via substitution [16] $\dfrac{\partial \mathbf{E}}{\partial t} \to \dfrac{\partial \mathbf{E}}{\partial t} + \dfrac{c}{n} \boldsymbol{\Omega} \times \mathbf{E}$, or, $\omega \to \omega + i\dfrac{c}{n} \boldsymbol{\Omega} \times \mathbf{E}$, in Eqs. (3.1)[2]. Neglecting higher-order terms proportional to $\Omega^2$ and $\dot{\boldsymbol{\Omega}}$, we arrive at
$$(\lambdabar_0^2 \nabla^2 + n^2) \mathbf{E} + 2in\lambdabar_0 \boldsymbol{\Omega} \times \mathbf{E} - \lambdabar_0^2 \nabla(\nabla \mathbf{E}) = 0. \qquad (3.6)$$
Here the second term describes the *Coriolis effect* caused by the rotation of the ray coordinate frame. This term is small: $\lambdabar_0 \Omega \sim \mu$, but should be taken into account in the first-order approximation in $\mu$.

Now we can project equation (3.6) onto the plane $(\mathbf{v}, \mathbf{w})$ orthogonal to the ray trajectory. For the projections of the last two terms we have in the first approximation in $\mu$:

---

[2] Here the wave velocity $c/n$ appears because we defined the angular velocity (3.4) with respect to the ray length $l$ rather than time. Alternatively, the Coriolis term arises upon the field differentiation along the ray: $\dot{\mathbf{E}} = \dfrac{d}{ds}(E_v \mathbf{v} + E_w \mathbf{w} + E_\parallel \mathbf{t}) = \dot{E}_v \mathbf{v} + \dot{E}_w \mathbf{w} + \dot{E}_\parallel \mathbf{t} + \boldsymbol{\Omega} \times \mathbf{E}$, where Eq. (3.5) was used.

$$\left[\bar{\lambda}_0^2\nabla(\nabla\mathbf{E})\right]_\perp \simeq 0, \; (\mathbf{\Omega}\times\mathbf{E})_\perp \simeq \Omega_\parallel(\mathbf{t}\times\mathbf{E}_\perp). \tag{3.7}$$

These equations take place owing to the wave transversality, Eqs. (3.2) and (3.3). Indeed, after substitution $-i\bar{\lambda}_0\nabla \to \mathbf{p}$ we have $\bar{\lambda}_0^2\nabla(\nabla\mathbf{E}) = -p^2 E_\parallel \mathbf{t}$, whereas in the Coriolis term we neglect small $E_\parallel \ll E$. As a result, the Maxwell equations for the transverse electric field $\mathbf{E}_\perp$ takes the form [16]:

$$\left(\bar{\lambda}_0^2\nabla^2 + n^2\right)\mathbf{E}_\perp + 2in\bar{\lambda}_0\Omega_\parallel(\mathbf{t}\times\mathbf{E}_\perp) = 0. \tag{3.8}$$

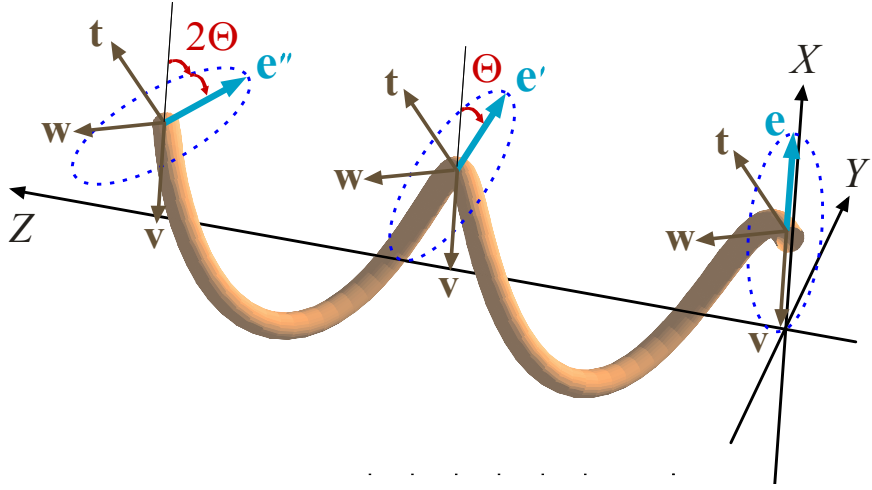

**Fig. 1.** Evolution of the polarization along a curved beam trajectory – a helix in this case. Polarization $\mathbf{e}$ and a ray-accompanying local coordinate frame $(\mathbf{v},\mathbf{w},\mathbf{t})$ are shown after each period of cyclical evolution of the tangent vector $\mathbf{t}$.

Equation (3.8) is a two-component vector equation which becomes diagonal in the basis of *circular* polarizations. Substituting the field as a superposition of right-hand and left-hand circular modes,

$$\mathbf{E}_\perp \propto E^+\boldsymbol{\xi} + E^-\boldsymbol{\xi}^*, \; \boldsymbol{\xi} = \frac{\mathbf{v}+i\mathbf{w}}{\sqrt{2}}, \; E^\pm = \frac{E_v \mp iE_w}{\sqrt{2}}, \tag{3.9}$$

we obtain from Eq. (3.8)

$$\left(\bar{\lambda}_0^2\nabla^2 + n^2\right)E^s + 2n\bar{\lambda}_0 s\Omega_\parallel E^s = 0, \; s = \pm 1. \tag{3.10}$$

Hereafter $s$ denotes the wave *helicity* indicating the two spin states of photons. According to Eq. (3.10), in the first approximation of the geometrical optics, these two states evolve independently, and the zero-order polarization degeneracy is lifted by the Coriolis term.

In the same approximation, Eq. (3.10) can be written as

$$\left[\bar{\lambda}_0^2\nabla^2 + \left(n + \bar{\lambda}_0 s\Omega_\parallel\right)^2\right]E^s = 0. \tag{3.10'}$$

Making the substitution $-i\bar{\lambda}_0\nabla \to \mathbf{p}$ and introducing Hamiltonian similarly to Eqs. (2.3) and (2.4), we arrive at the Hamiltonian

$$\mathcal{H} = p - n - \bar{\lambda}_0 s\Omega_\parallel = 0. \tag{3.11}$$

Despite the above 'quantum-to-classical' substitution, the Hamiltonian (3.11) includes $\bar{\lambda}$-order correction from the Coriolis term. The corresponding Lagrangian of the problem is

$$\mathcal{L} = \mathcal{L}_0 + s\mathcal{L}_1, \; \mathcal{L}_1 = \bar{\lambda}_0\Omega_\parallel, \tag{3.12}$$

where $\mathcal{L}_0 = n - p + \mathbf{p}\dot{\mathbf{r}}$ is the scalar Lagrangian (2.6), whereas $s\mathcal{L}_1$ is the Lagrangian describing the *spin-orbit coupling* of light. Similar spin-orbit terms have appeared before in the theory of spinning particles, see [29,32,60–63].

The spin-orbit Lagrangian is proportional to the longitudinal component of the angular velocity of the co-moving coordinate frame, which is

$$\Omega_\| = -\mathbf{v}\dot{\mathbf{w}} = i\xi^*\dot{\xi}. \tag{3.13}$$

If we introduce the spin angular momentum of the circularly polarized wave as $\mathbf{\Sigma} = \lambdabar_0 s\mathbf{t}$ (for massless particles spin is directed along the momentum), the spin-orbit Lagrangian equals

$$_s\mathcal{L}_1 = \mathbf{\Sigma}\mathbf{\Omega}. \tag{3.14}$$

This Lagrangian has the form of the Coriolis term in systems with intrinsic angular momentum [64–67]. At the same time, Eq. (3.14) shows that it can be treated as the *angular Doppler effect* term, which describes the phase shift of the wave carrying intrinsic angular momentum $\mathbf{\Sigma}$ and undergoing rotational evolution with angular frequency $\mathbf{\Omega}$ [68–80].[3] Note that here $\mathbf{\Sigma}\mathbf{\Omega} > 0$ corresponds to a red shift of the wave frequency, because the medium rotates with respect to the chosen coordinate frame with angular velocity $-\mathbf{\Omega}$.

## 3.3. Berry connection and curvature

The spin-orbit Lagrangian (3.12) brings about an additional polarization-dependent wave phase. In order to describe this phase, we first introduce new representation and quantities which are important for what follows.

The co-moving coordinate frame $(\mathbf{v}, \mathbf{w}, \mathbf{t})$ is attached to the direction of the wave momentum $\mathbf{p}$, and the polarization evolution of the wave is essentially *momentum*-dependent (rotations of the ray coordinate frame are independent of the particular space coordinates, $\mathbf{r}$). Therefore, one can parameterize basis vectors of the ray coordinate frame by the momentum:

$$\mathbf{t} = \mathbf{t}(\mathbf{p}),\ \mathbf{v} = \mathbf{v}(\mathbf{p}),\ \mathbf{w} = \mathbf{w}(\mathbf{p}),\ \text{or,}\ \xi = \xi(\mathbf{p}). \tag{3.15}$$

Transition from the $l$-parameterization to the $\mathbf{p}$-parameterization is performed via the substitution $\dfrac{d}{dl} \to \dfrac{d\mathbf{p}}{dl}\dfrac{\partial}{\partial \mathbf{p}}$, and the spin-orbit Lagrangian (3.12)–(3.14) takes the form

$$\mathcal{L}_1 = -\lambdabar_0 \mathbf{A}(\mathbf{p})\dot{\mathbf{p}}, \tag{3.16}$$

where

$$A_i = \mathbf{v}\frac{\partial \mathbf{w}}{\partial p_i} = -i\xi^* \frac{\partial \xi}{\partial p_i} \tag{3.17}$$

is the so-called *Berry connection* or *Berry gauge field*. In such representation, the spin-orbit interaction of light, Eq. (3.16), acquires the same form as the spin-orbit interaction of electrons stemming from the Dirac equation [29,32,54,81–83].

As we will see, the Berry connection relates the wave polarization $\mathbf{e} = \mathbf{E}_\perp / E_\perp$ in the neighboring points $\mathbf{p}$ and $\mathbf{p} + d\mathbf{p}$ of momentum space. Since the polarization depends only on the *direction* of momentum, $\mathbf{t} = \mathbf{p}/p$, the evolution in the $\mathbf{p}$ space can be projected onto the unit sphere $S^2 = \{\mathbf{t}\}$. In this manner, polarization vector $\mathbf{e}$ is tangent to this sphere, and the Berry connection determines the natural *parallel transport* of $\mathbf{e}$ over the $S^2$ sphere, Fig. 2 [1–7,9,13–15,44,84–90].

The curvature tensor corresponding to the Berry connection is

$$F_{ij} = \frac{\partial A_j}{\partial p_i} - \frac{\partial A_i}{\partial p_j}, \tag{3.18}$$

---

[3] Usually, the angular Doppler shift is considered for a wave interacting with a rotating element. If this element changes only the polarization state, the shift can be associated with the Pancharatnam–Berry phase. When it changes only the propagation direction of the wave, the shift can be regarded as a manifestation of the spin-redirection Rytov−Vladimirskii−Berry phase considered here. Thus, the angular Doppler shift links and unifies the two types of the geometric phase in optics [67].

which yields

$$F_{ij} = \frac{\partial \mathbf{v}}{\partial p_i}\frac{\partial \mathbf{w}}{\partial p_j} - \frac{\partial \mathbf{v}}{\partial p_j}\frac{\partial \mathbf{w}}{\partial p_i} = -i\left(\frac{\partial \xi^*}{\partial p_i}\frac{\partial \xi}{\partial p_j} - \frac{\partial \xi^*}{\partial p_j}\frac{\partial \xi}{\partial p_i}\right). \quad (3.19)$$

This is an antisymmetric tensor characterized by the dual vector $\mathbf{F}$: $F_{ij} = \varepsilon_{ijk}F_k$, so that:

$$\mathbf{F} = \frac{\partial}{\partial \mathbf{p}} \times \mathbf{A}, \quad (3.18')$$

This the *Berry curvature* or *Berry field strength*. Direct calculations (see Appendix) show that, independently on the choice of the coordinate frame $(\mathbf{v}, \mathbf{w}, \mathbf{t})$, the Berry curvature is equal to

$$\mathbf{F} = \frac{\mathbf{p}}{p^3}. \quad (3.20)$$

On the surface of the unit $\mathbf{t}$-sphere (i.e., at $p = 1$), the Berry curvature equals $\mathbf{F} = \mathbf{t}$ indicating the unit Gaussian curvature of the surface.

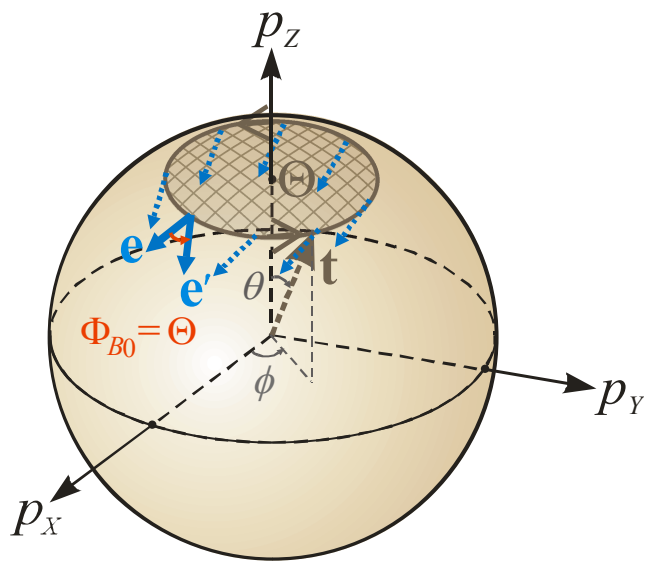

**Fig. 2.** Scheme of the parallel transport of the polarization vector $\mathbf{e}$ over the unit sphere $S^2 = \{\mathbf{t}\}$ in momentum space. Trajectory of evolution of the $\mathbf{t}$ vector and polarization variations correspond to one period of the helical trajectory depicted in Fig. 1.

Geometrical properties of the Berry connection and curvature have been discussed in numerous reviews and papers [1–7,9,13–15,44,84–90]. Important to note that these quantities allow also *dynamical* interpretation. Indeed, the Berry connection $\mathbf{A}$ appears in Lagrangian (3.16) as an external vector-potential affecting the evolution of light.[4] In turn, the Berry curvature $\mathbf{F}$, Eq. (3.18), plays the role of the effective 'magnetic field' corresponding to the vector-potential $\mathbf{A}$. Remarkably, this field takes the form of a '*magnetic monopole*' located at the origin of $\mathbf{p}$-space [12,17–19,26–35] (cf. [38,39]). As it will be shown, the topological monopole affects the wave motion as a rather real physical entity (see Section 3.5) [35].

The gauge properties of the potential $\mathbf{A}$ and field $\mathbf{F}$ are closely related to the choice of the co-moving frame $(\mathbf{v}, \mathbf{w}, \mathbf{t})$. The Berry connection and curvature are defined through the basis vectors $(\mathbf{v}, \mathbf{w})$, or $\xi$, which, in turn, are defined up to an arbitrary rotation about $\mathbf{t}$. Such a local rotation of the coordinate frame on an angle $\alpha = \alpha(\mathbf{p})$, induces transformation of the basis vectors $(\mathbf{v}, \mathbf{w})$:

$$\begin{pmatrix} \mathbf{v} \\ \mathbf{w} \end{pmatrix} \to \begin{pmatrix} \cos\alpha & \sin\alpha \\ -\sin\alpha & \cos\alpha \end{pmatrix} \begin{pmatrix} \mathbf{v} \\ \mathbf{w} \end{pmatrix} \tag{3.21}$$

---

[4] Compare Lagrangian (3.16) with the Lagrangian of electric charge $q$ in an external electromagnetic vector-potential $\mathcal{A}(\mathbf{r})$: $\mathcal{L}_q = (q/c)\mathcal{A}(\mathbf{r})\dot{\mathbf{r}}$ [91].

For the basis vectors of circular polarizations, $\xi$, Eq. (3.8), the transformation (3.21) results in

$$\xi \to \exp(-i\alpha)\xi, \tag{3.22}$$

i.e., $SO(2)$ rotation of $(\mathbf{v}, \mathbf{w})$ is equivalent to $U(1)$ gauge transformation of $\xi$ [44]. From Eq. (3.17), it is seen that the gauge transformation (3.22) generates the following change of the Berry connection:

$$\mathbf{A} \to \mathbf{A} - \frac{\partial \alpha}{\partial \mathbf{p}}. \tag{3.23}$$

At the same time, this transformation (akin to the gradient gauge transformation of the electromagnetic vector-potential) does not influence the Berry curvature:

$$\mathbf{F} \to \mathbf{F}. \tag{3.24}$$

Therefore, all physical quantities which are independent of the choice of coordinate frame (e.g., ray trajectories), must be dependent on the Berry curvature $\mathbf{F}$ rather than on the gauge-variant connection $\mathbf{A}$.

## 3.4. Berry phase and polarization evolution

The wave phase is obtained by integration of the Lagrangian (3.12) along the ray. In addition to the scalar phase (2.7), it acquires a polarization-dependent phase from the spin-orbit Lagrangian:

$$\Phi = \Phi_0 - s\Phi_B, \quad \Phi_B = -\lambdabar_0^{-1} \int \mathcal{L}_1 dl = -\int \Omega_\parallel dl. \tag{3.25}$$

Substituting Eqs. (3.13) and (3.16) we can write this phase as

$$\Phi_B = \int \mathbf{v}\, d\mathbf{w} = -i \int \xi^* d\xi = \int_\Gamma \mathbf{A}\, d\mathbf{p}, \tag{3.26}$$

where $\Gamma$ is the contour of the wave evolution in the $\mathbf{p}$ space. Equation (3.26) describes the *spin-redirection Berry geometric phase* acquired by the right-hand and left-hand circularly polarized modes with the opposite signs [4–20].[5] The Berry phase is a manifestation of the Coriolis or angular Doppler effect originated from the rotation of the wave bearing spin angular momentum.

The Berry phase determines variations of the polarization of light along the zero-approximation ray (2.5). Let us introduce the unit complex two-component Jones vector in the basis of circular polarizations:

$$|\psi\rangle = \begin{pmatrix} e^+ \\ e^- \end{pmatrix}, \quad \langle\psi|\psi\rangle = 1, \tag{3.27}$$

where $e^\pm = E^\pm / E_\perp$. The Berry phases (3.25) and (3.26) lead to the evolution of the Jones vector as follows:

$$|\psi(l)\rangle = \begin{pmatrix} \exp(-i\Phi_B) & 0 \\ 0 & \exp(+i\Phi_B) \end{pmatrix} |\psi(0)\rangle. \tag{3.28}$$

This phase difference of the circular polarizations is equivalent to the rotation of the linear polarizations on the angle $\Phi_B$.[6] In the generic case of an elliptical polarization, Eq. (3.28) describes the same turn of the polarization ellipse on angle $\Phi_B$ with its eccentricity conserved, Fig. 1 [4–20,59]. Such rotation of the polarization along a curved ray trajectory had been described by Rytov and Vladimirskii long before the Berry phase has been discovered [4,8,9].

---

[5] It should be noted that we refer to papers dealing with two types of optical systems: (i) waves propagating freely in a gradient-index medium and (ii) waves propagating in a curved single-mode dielectric waveguide. Rigorously speaking, the geometrical optics (ray) description is inapplicable in single-mode waveguides [14], but the polarization evolution and Berry phase of the waveguide modes obey the same geometrical formalism.

The differential form of Eq. (3.28) is [27,34]:

$$|\dot{\psi}\rangle = i\Omega_\| \hat{\sigma}_3 |\psi\rangle = -i(\mathbf{A}\dot{\mathbf{p}})\hat{\sigma}_3 |\psi\rangle. \tag{3.29}$$

Hereafter we use the Pauli matrices

$$\hat{\sigma}_1 = \begin{pmatrix} 0 & 1 \\ 1 & 0 \end{pmatrix}, \ \hat{\sigma}_2 = \begin{pmatrix} 0 & -i \\ i & 0 \end{pmatrix}, \ \hat{\sigma}_3 = \begin{pmatrix} 1 & 0 \\ 0 & -1 \end{pmatrix}. \tag{3.30}$$

Equation (3.29) is similar to the equation for polarization evolution in quantum spin systems [54,92,93].

From Eq. (3.29) it follows that polarization measured in the ray-accompanying coordinate frame $(\mathbf{v}, \mathbf{w}, \mathbf{t})$ rotates with the local angular velocity $-\Omega_\| \mathbf{t}$. Hence, the Berry phase describes a sort of *inertia* of the electric field which remains *locally* non-rotating about the ray in the laboratory frame. This is seen from the dynamics of the unit polarization vector

$$\mathbf{e} = \frac{\mathbf{E}_\perp}{E_\perp} = e^+ \boldsymbol{\xi} + e^- \boldsymbol{\xi}^*. \tag{3.31}$$

Differentiating Eq. (3.31) and substituting derivatives of $\boldsymbol{\xi}$ and $e^\pm$ from Eqs. (3.4), (3.5), (3.9), (3.27), and (3.29), we arrive at [4]

$$\dot{\mathbf{e}} = -(\mathbf{e}\dot{\mathbf{t}})\mathbf{t}. \tag{3.32}$$

This is a well-known equation for the *parallel transport* of vector $\mathbf{e}$ along the ray. According to it, $\mathbf{e}$ does not experience local rotation about $\mathbf{t}$. In contrast to the previous Eqs. (3.25)–(3.29), equation (3.32) is *independent* of the local coordinate frame $(\mathbf{v}, \mathbf{w}, \mathbf{t})$ and describes the polarization dynamics in a universal way.

The parallel transport of vector $\mathbf{e}$ can be equally considered either along the ray trajectory or over the $\mathbf{t}$-sphere, Figs. 1 and 2. Equation (3.32) is *non-integrable*, i.e., its solution $\mathbf{e}(l)$ is *non-local* due to the non-integrability of the Berry phase (3.25) and (3.26). Therefore, despite the local non-rotational evolution, *globally* the polarization $\mathbf{e}$ can turn after a cyclical evolution of the $\mathbf{t}$ vector, Figs. 1 and 2. Indeed, a series of local rotations of the $\mathbf{t}$-vector about different axes may produce a turn of $\mathbf{e}$ about the $\mathbf{t}$-vector itself. This is a result of non-commutativity of $SO(3)$ rotations in space [7]. The angle of the resulting rotation of $\mathbf{e}$ equals the Berry phase $\Phi_B$.

For a cyclic evolution in momentum space (which corresponds to a ray trajectory with the coincident input and output directions of propagation), the Berry phase (3.26) equals [4–20]

$$\Phi_{B0} = \oint_\Gamma \mathbf{A}\,d\mathbf{p} = \iint_\Pi \mathbf{F}\,d^2\mathbf{p} = \Theta. \tag{3.33}$$

Here $\Pi$ is a surface spanned over the closed contour $\Gamma$ (so that $\Gamma = \partial\Pi$), and $\Theta$ is the solid angle cut by $\Gamma$, i.e., the area enclosed by the projection of the loop $\Gamma$ onto the $\mathbf{t}$ sphere, Fig. 2. The last equality in Eq. (3.33) is obtained using Eq. (3.20). Equation (3.33) indicates the turn of the vector $\mathbf{e}$ parallel-transported over the $\mathbf{t}$ sphere. At the same time, it shows that the Berry phase is analogous to the Aharonov–Bohm (or Dirac) phase in quantum mechanics [94–96]. It is equal to the contour integral of the effective vector-potential **A** or to the flux of the 'magnetic field' **F** through this contour.

---

[6] Transition to the basis of linear polarizations, $e_{v,w} = E_{v,w}/E_\perp$, is realized via substitution $\begin{pmatrix} e^+ \\ e^- \end{pmatrix} = \frac{1}{\sqrt{2}} \begin{pmatrix} 1 & -i \\ 1 & i \end{pmatrix} \begin{pmatrix} e_v \\ e_w \end{pmatrix}$.

The solid angle $\Theta$ can be calculated as

$$\Theta = \oint_{\tilde{\Gamma}} (1 - \cos\theta) d\phi = \iint_{\tilde{\Pi}} \sin\theta \, d\theta d\phi. \qquad (3.34)$$

Here $\tilde{\Gamma}$ and $\tilde{\Pi}$ are the projections of $\Gamma$ and $\Pi$ onto the **t** sphere, whereas $(\theta, \phi)$ are the spherical angles on the **t** sphere, which specify direction of the wave momentum with respect to the laboratory coordinate frame $(XYZ)$: $\mathbf{p} \equiv (p_X, p_Y, p_Z) = p(\sin\theta\cos\phi, \sin\theta\sin\phi, \cos\theta)$, Fig. 2. Equations (3.33) and (3.34) imply that the Berry connection and curvature can be written in spherical coordinates $(p, \theta, \phi)$ as (see Section A.1)

$$\mathbf{A} \equiv (A_p, A_\theta, A_\phi) = p^{-1}(0, 0, \sin^{-1}\theta - \tan^{-1}\theta), \quad \mathbf{F} \equiv (F_p, F_\theta, F_\phi) = -p^{-2}(1, 0, 0). \qquad (3.35)$$

The Berry connection is determined up to a gauge transformation (3.23) which does not change the contour integral (3.33), and the Berry curvature is the monopole (3.20).

The polarization rotation corresponding to the Berry phase (3.33) and (3.34) has been measured experimentally in single-mode optical fibers coiled in a helix similar to Fig. 1 [10,11] and for helical rays in a multimode rectilinear waveguide [20,35], cf. Section 4.

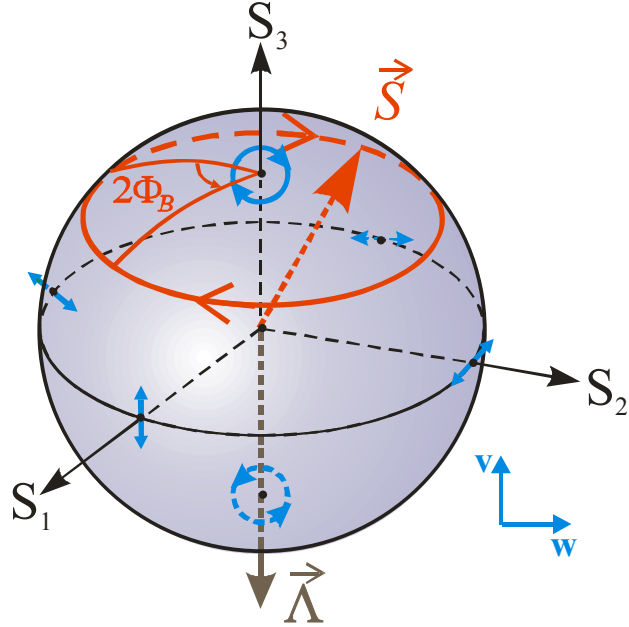

**Fig. 3.** Representation of the polarization evolution along a helical trajectory of Fig. 1 on the Poincaré sphere. The reference circular and linear polarization states are depicted on the sphere with respect to the $(\mathbf{v}, \mathbf{w})$ frame.

We have written the equation of motion describing the polarization dynamics in the Jones representation, Eq. (3.29). An alternative formalism, frequently used in polarization optics, is the Stokes-vector representation on the Poincaré sphere. To address this representation, let us exploit the analogy between the optics and quantum mechanics. The vector of Pauli matrices, $\hat{\vec{\sigma}} = (\hat{\sigma}_1, \hat{\sigma}_2, \hat{\sigma}_3)$, can be treated in Eq. (3.29) as a spin operator. Then, the expectation value of the spin vector is given by

$$\vec{S} = \langle \psi | \hat{\vec{\sigma}} | \psi \rangle, \quad \vec{S}^2 = 1. \qquad (3.36)$$

The unit vector $\vec{S}$ is the Stokes vector representing the wave polarization on the Poincaré sphere – the north and south poles represent right- and left-hand circular polarizations, whereas the equator represents linear polarizations, Fig. 3. By differentiating expression (3.36) and using Eq. (3.29) we find that the Stokes vector obeys the following precession equation [34,35]:

$$\dot{\vec{S}} = \vec{\Lambda} \times \vec{S}, \quad \vec{\Lambda} = -2\vec{\Omega}_\parallel = 2(\mathbf{A}\dot{\mathbf{p}})\vec{u}_3, \tag{3.37}$$

where $\vec{u}_3$ is the basis vector of the $S_3$ axis. Thus, upon the propagation of light along the ray trajectory, the Stokes vector precesses about the $S_3$ axis on the Poincaré sphere with the angular velocity $\vec{\Lambda} = -2\vec{\Omega}_\parallel$, Fig. 3. A cyclical evolution on the $\mathbf{t}$ sphere, Fig. 2, corresponds to the rotation on the angle $2\Phi_{B0}$ on the Poincaré sphere.[7]

The Stokes vector can be interpreted as a *pseudospin* of photons. Its third component, $S_3 = |e^+|^2 - |e^-|^2 \in [-1,1]$, indicates the wave helicity and represents a continuous analogue of the quantized helicity $s = \pm 1$. In virtue of Eq. (3.37), the helicity is conserved upon the light propagation:

$$\dot{S}_3 = 0. \tag{3.38}$$

This evidences *adiabatic* regime of the wave evolution [4,12,14,17], i.e., independence of the right- and left-hand circular polarizations. It is worth remarking that the pseudospin $\vec{S}$ completely describes the polarization state of the wave, while the real spin angular momentum $\mathbf{\Sigma} = \hbar_0 s \mathbf{t}$ does not [97]. The Jones and Stokes representations characterize the polarization state of light, respectively, by the complex two-component vector $|\psi\rangle$ and the real three-component vector $\vec{S}$. In the generic case, they obey $SU(2)$ and $SO(3)$ evolutions and correspond, respectively, to the Schrödinger and Heisenberg representations of spin in quantum mechanics [34,98].

Finally, note that polarization evolution equations (3.29) and (3.37) are not gauge-invariant with respect to transformations (3.23) because the field components and, hence, the polarization state depend on the choice of the coordinate frame $(\mathbf{v}, \mathbf{w}, \mathbf{t})$.

## 3.5. Ray equations and spin Hall effect

We have considered the phase effects induced by the spin-orbit coupling of light. Here we show that it also perturbs the ray trajectories. The total wave Lagrangian can be written from Eqs. (2.4), (2.6), (3.12) and (3.16) as

$$\mathcal{L}(\mathbf{r}, \dot{\mathbf{r}}, \mathbf{p}, \dot{\mathbf{p}}) = n - p + \mathbf{p}\dot{\mathbf{r}} - \hbar_0 s \mathbf{A} \dot{\mathbf{p}}. \tag{3.39}$$

The Euler–Lagrange equations for the Lagrangian (3.39) varying independently with respect to $\mathbf{p}$ and $\mathbf{r}$ result in [24–35]

$$\dot{\mathbf{p}} = \nabla n, \quad \dot{\mathbf{r}} = \frac{\mathbf{p}}{p} + \hbar_0 s \dot{\mathbf{p}} \times \mathbf{F} = \frac{\mathbf{p}}{p} + \hbar_0 s \frac{\dot{\mathbf{p}} \times \mathbf{p}}{p^3}. \tag{3.40}$$

These are the ray equations of the first-order approximation of the geometrical optics. As compared to the zero-order equations (2.5), Eqs. (3.40) acquire a new, topological term proportional to the wave helicity $s$. This term describes a polarization-dependent *transverse deflection* of the ray trajectory, which is orthogonal both to the wave momentum $\mathbf{p}$ and inhomogeneity gradient $\dot{\mathbf{p}} = \nabla n$. Thus, light beams with different polarizations propagate along slightly different trajectories, Fig. 4. This striking effect is known as the *spin-Hall effect of light* or *optical Magnus effect* [24–35]. It can be treated as an effective circular *birefringence* of the inhomogeneous medium [25,26], but, in fact, it arises because of the intrinsic property (spin) of light coupled to a curved ray trajectory. The topological term in Eqs. (3.40) makes the velocity and momentum non-collinear: $\dot{\mathbf{r}} \not\parallel \mathbf{p}$ (i.e., the wave vector is not tangent to the ray), which is also characteristic for anisotropic media [59].

---
[7] The factor of 2 appears here because one complete $2\pi$ turn of the polarization ellipse in the real space corresponds to a $4\pi$ double-turn on the Poincaré sphere.

Equations (3.40) depend on the Berry curvature **F** and therefore are gauge-invariant, i.e., independent on the coordinate frame. In the spin Hall effect, the Berry curvature manifests itself *dynamically* – the polarization term in Eq. (3.40) represents a '*Lorentz force*' from the 'magnetic monopole' (3.20). Since the field **F** appears in the momentum **p** space, the 'force' $\dot{\mathbf{p}} \times \mathbf{F}$ occurs in the equation for $\dot{\mathbf{r}}$. Formally speaking, this is a correction to the velocity, the so-called *anomalous velocity* rather than force, cf. [37–43]. Despite a long history of the geometrical optics method, the polarization-dependent perturbation of the ray equations was introduced only recently. The point is that, traditionally, the ray trajectories were defined in the 'classical' limit, $\lambdabar_0 \to 0$ [59]. The spin-orbit term in Eqs. (3.40) is proportional to $\lambdabar_0$, and, thus, can be treated as a '*semiclassical*' correction to the ray equations.

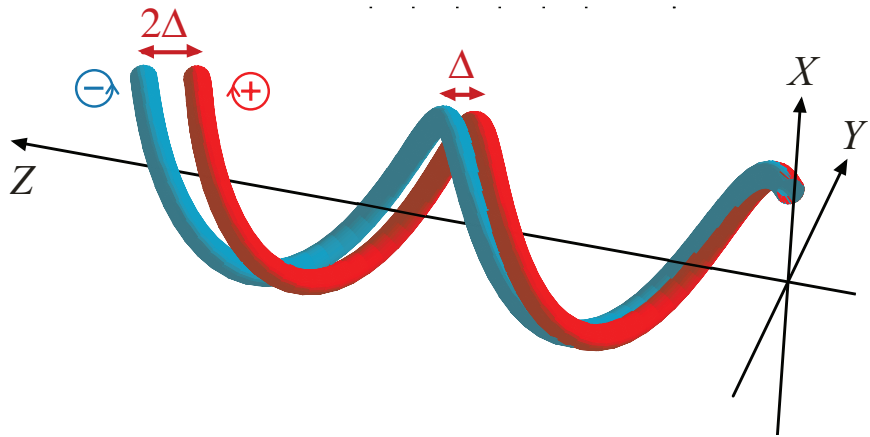

**Fig. 4.** The spin Hall effect of light: deflections of the trajectories of the light propagation for right-hand (+) and left-hand (−) polarizations, for a helical path from Fig. 1. The splitting of the rays per one period of the helix is $\Delta = 2|\delta \mathbf{r}_0|$.

Ray equations (3.40) describe the trajectory of the *center of gravity* of a polarized wave beam or wave packet[8]. They are written for circularly-polarized eigenmodes with $s = \pm 1$. To determine motion of the center of gravity of a beam with an elliptical polarization, one has to change the discrete helicity $s$ into its continuous counterpart:

$$s \to S_3 = \langle \psi | \hat{\sigma}_3 | \psi \rangle = |e^+|^2 - |e^-|^2. \quad (3.41)$$

It should be noted, however, that an elliptically polarized beam will split upon the propagation into two overlapping but mutually displaced circularly polarized beams which propagate according to Eqs. (3.40). The ray equations (3.40) with (3.41) (which depend on $\vec{S}$ or $|\psi\rangle$) and equation for polarization dynamics (3.29) or (3.37) (which depends on **p** and **r**) form a complete set of the equations of motion. The intrinsic and extrinsic degrees of freedom are coupled in these equations due to the spin-orbit interaction of light.

The spin Hall effect of light is a dynamical manifestation of the Coriolis effect. The transverse deflection occurs upon a bending of the trajectory of light carrying intrinsic angular momentum. Indeed, the spin-dependent term of Eqs. (3.40) is proportional to the curvature of the ray and directed orthogonally to the local propagation plane (see Section A.2). The spin Hall effect is also closely related to the conservation of the total angular momentum of light. Indeed, in a spherically-symmetric medium, $n(\mathbf{r}) = n(r)$, the ray equations (3.40) possess the following integral of motion [27]:

$$\mathbf{J} = \mathbf{L} + \mathbf{\Sigma} = \mathbf{r} \times \mathbf{p} + \bar{\lambda}_0 s \mathbf{t} = \text{const}. \quad (3.42)$$

---

[8] We imply Gaussian-type wave beams or packets which do not bear an intrinsic orbital angular momentum.

Here $\mathbf{J}$ is obviously associated with the total angular momentum of light, consisting of the extrinsic orbital angular momentum $\mathbf{L} = \mathbf{r} \times \mathbf{p}$ and the intrinsic spin angular momentum $\mathbf{\Sigma}$. It is the spin-orbit term in Eqs. (3.40) that provides for the conservation of $\mathbf{J}$: variations in the direction of $\mathbf{\Sigma}$ caused by the wave refraction are compensated by the transverse shift of the ray which changes the orbital part $\mathbf{L}$. In a medium, with a cylindrical symmetry with respect to the $Z$-axis, the conserved quantity is $J_z$.

The conservation of the total angular momentum, Eq. (3.42), links the spin Hall effect of light to another remarkable phenomenon – the transverse Imbert-Fedorov shift [24,27,99–118]. The latter effect is a similar polarization-dependent displacement of a wave beam reflected or transmitted at a plane dielectric interface. In fact, the spin Hall effect in a gradient-index medium and the Imbert-Fedorov shift represent, respectively, the weak- and strong-scattering limits of the polarization transport induced by the spin-orbit interaction of light. In the limiting case of low contrast between the two media (the ray trajectory is changed weakly in this case), the expression for the Imbert-Fedorov shift corresponds to the spin Hall effect from Eqs. (3.40) [24,119]. However, in the general case, the Imbert-Fedorov shift cannot be described within the Berry-phase formalism, because of the non-adiabatic evolution of light upon the reflection or transmission at a sharp interface. Despite a long history of theoretical and experimental studies, the correct generic expressions for the Imbert-Fedorov shift [115,116] and accurate experimental measurements [117] were achieved only recently.

The spin-dependent perturbations of the ray trajectories, $\delta \mathbf{p}$ and $\delta \mathbf{r}$, can be found from Eqs. (3.40) using substitution

$$\mathbf{p} = \mathbf{p}^{(0)} + \delta \mathbf{p}, \quad \mathbf{r} = \mathbf{r}^{(0)} + \delta \mathbf{r}, \quad (3.43)$$

where $\mathbf{p}^{(0)}$ and $\mathbf{r}^{(0)}$ satisfy the scalar equations (2.5). In this manner, we obtain in the first-order perturbation with respect to $\delta \mathbf{p}$ and $\delta \mathbf{r}$:

$$\delta \dot{\mathbf{p}} = (\delta \mathbf{r} \nabla) \nabla n, \quad \delta \dot{\mathbf{r}} = \frac{\delta \mathbf{p}_\perp}{p^{(0)}} + \bar{\lambda}_0 s \frac{\dot{\mathbf{p}}^{(0)} \times \mathbf{p}^{(0)}}{p^{(0)3}}, \quad (3.44)$$

where $\delta \mathbf{p}_\perp = \delta \mathbf{p} - (\mathbf{t} \delta \mathbf{p}) \mathbf{t}$, $\mathbf{t} = \mathbf{p}^{(0)} / p^{(0)}$. In the important special case $\delta \mathbf{p} \equiv 0$, Eqs. (3.44) are simplified to

$$\delta \dot{\mathbf{r}} = \bar{\lambda}_0 s \frac{\dot{\mathbf{p}}^{(0)} \times \mathbf{p}^{(0)}}{p^{(0)3}}. \quad (3.45)$$

From here, the displacement of the ray trajectory is [25,26,31,34]:

$$\delta \mathbf{r} = -\bar{\lambda}_0 s \int_\Gamma \frac{\mathbf{p} \times d\mathbf{p}}{p^3} = -\bar{\lambda}_0 s \int_\Gamma \mathbf{F} \times d\mathbf{p}, \quad (3.46)$$

Thus, akin to the Berry phase (3.26) and (3.33), the ray displacement takes the form of a non-local contour integral in momentum space, which can grow unlimitedly with the length of the trajectory.

The magnitude of the displacement $\delta \mathbf{r}$ can be estimated from Eq. (3.46) as the wavelength $\lambda$ for a complete $2\pi$ turn of the direction of propagation (e.g., for a circular trajectory) [25]. Typically, this effect is much smaller than the width of a paraxial optical beam, but nonetheless it is observable, since the position of center of gravity of the beam can be measured with a sub-wavelength accuracy. Furthermore, the displacement is accumulated for repeating, cyclical trajectories in the $\mathbf{p}$ space – e.g., for circular or helical trajectories in the real space, Fig. 4 [25,35].

Recently, the spin Hall effect of light and the ray equations (3.40) were verified by direct measurements for a helical trajectories as in Fig. 4 [35]. There had been a number of remarkable experiment [120–123] which observed the spin-orbit coupling and polarization transport of light in complex speckle fields of multimode optical fibers (see also [24]), but all of them (apart from numerical simulations in [123]) dealt with step-index index fibers, so that these results should rather be associated with the Imbert-Fedorov shifts upon the reflection at sharp inhomogeneities. In addition, propagation of light in optical fibers does not allow one to observe the geometrical-optics ray trajectories hidden in the complex speckle patterns of the interfering modes.

## 4. Example: Berry phase and spin Hall effect for a helical ray

A characteristic example that exhibits both the Berry phase and the spin Hall effect, is the light propagation along a helical ray in a medium with cylindrical symmetry.

*4.1. Helical rays in the scalar approximation*

Let the medium be characterized by the refractive index $n = n(R)$ in the laboratory cylindrical coordinates $(R, \Phi, Z)$ used instead of the Cartesian coordinates $(X, Y, Z)$. In the cylindrical coordinates, the scalar-approximation ray equations (2.5) yield [59]

$$\dot{R} = \frac{p_R}{n}, \quad R\dot{\Phi} = \frac{p_\Phi}{n}, \quad \dot{Z} = \frac{p_Z}{n}; \quad \dot{p}_R = n' + \frac{p_\Phi^2}{nR}, \quad \dot{p}_\Phi = -\frac{p_R p_\Phi}{nR}, \quad \dot{p}_Z = 0, \quad (4.1)$$

where $n' \equiv dn/dR$. The generic helical ray trajectory can be written as

$$R = \rho, \quad \rho\Phi = l\sin\vartheta, \quad Z = l\cos\vartheta,$$
$$p_R = 0, \quad p_\Phi = n\sin\vartheta, \quad p_Z = n\cos\vartheta, \quad (4.2)$$

where $n = n(\rho)$, and the helix is characterized by two parameters: the radius $\rho$ and the angle $\vartheta$ between the tangent vector and the $Z$ axis. Ray trajectory (4.2) satisfies Eqs. (4.1) when

$$\sin^2\vartheta = -\left.\frac{\rho n'}{n}\right|_{R=\rho}. \quad (4.3)$$

Equations (4.2) and (4.3) determine the scalar-approximation helical ray trajectories in the cylindrical medium.

*4.2. Ray coordinate frames*

To describe the polarization variations along the trajectory, one needs to introduce one of the specific ray coordinate frames considered in the Appendix. The tangent vector $\mathbf{t} = \mathbf{p}/p$ is determined by the ray trajectory (4.2):

$$\mathbf{t} = \sin\vartheta \, \mathbf{e}_\Phi + \cos\vartheta \, \mathbf{e}_Z. \quad (4.4)$$

Following Section A.1, the tangent $\mathbf{t}$ is given by spherical angles $(\theta, \phi) = \left(\vartheta, \frac{\pi}{2} + \varphi\right)$. Substituting this in Eqs. (A3) and introducing the basis vectors of the cylindrical laboratory coordinates,

$$\mathbf{e}_X = \mathbf{e}_R \cos\Phi - \mathbf{e}_\Phi \sin\Phi, \quad \mathbf{e}_Y = \mathbf{e}_R \sin\Phi + \mathbf{e}_\Phi \cos\Phi,$$

we obtain

$$\mathbf{v} = -\mathbf{e}_R, \quad \mathbf{w} = -\mathbf{e}_\Phi \cos\vartheta + \mathbf{e}_Z \sin\vartheta. \quad (4.5)$$

This specifies the Euler-angles ray coordinate frame with $\mathbf{v}$ directed along the radial coordinate, see Fig. 1.

Alternatively, following Section A.2, one can determine the Frenet basis along the ray. The tangent **t** is constant in the cylindrical coordinate, but it evolves in the Cartesian frame. From ray equations (2.5), easy to see that the derivative $\dot{\mathbf{t}}$ is directed along $\nabla_\perp n \equiv \nabla n - (\mathbf{t}\nabla n)\mathbf{t}$, i.e., along $-\mathbf{e}_R$ in our case. Then, Eqs. (A8) with Eq. (4.4) yield the principal normal and binormal to the ray:

$$\mathbf{n} = -\mathbf{e}_R, \quad \mathbf{b} = -\mathbf{e}_\Phi \cos\vartheta + \mathbf{e}_Z \sin\vartheta. \tag{4.6}$$

Thus, in the case of a helical trajectory, the Frenet coordinate frame (4.6) coincides with the Euler-angles one, Eqs. (4.5): $(\mathbf{n},\mathbf{b}) \equiv (\mathbf{v},\mathbf{w})$. They represent a natural choice for the description of the polarization evolution along the ray.

The parallel-transport frame can be constructed using Eq. (A14) and the Berry phase calculated below.

## 4.3. Evolution of polarization and the Berry phase

Since the Euler-angles and Frenet frames coincide with each other, the Berry-phase expressions are equivalent as well. The torsion of the helix (4.2) equals $T^{-1} = \sin 2\vartheta/(2\rho)$, and from Eqs. (3.12), (3.16), (A4), and (A10) we have:

$$-\Omega_\| = \mathbf{A}\dot{\mathbf{p}} = -\dot{\phi}\cos\theta = -T^{-1} = -\frac{\sin 2\vartheta}{2\rho}. \tag{4.7}$$

This quantity is constant and its integration results in the Berry phase

$$\Phi_B = -\frac{\sin 2\vartheta}{2\rho}l \tag{4.8}$$

and the polarization evolution in the Jones representation, Eq. (3.28). Thus, the Berry phase grows linearly with the propagation distance $l$, and polarization ellipse experience a uniform rotation with angular velocity (4.7) in the ray reference frame, see Fig. 1.

In the Stokes-vector formalism, Eq. (3.37), the Stokes vector precesses uniformly with angular velocity, Fig. 3:

$$\vec{\Lambda} = -\frac{\sin 2\vartheta}{\rho}\vec{u}_3. \tag{4.9}$$

Integration of Eq. (3.37) with $\vec{\Lambda} = \text{const}$ yields

$$S_1 = S_{10}\cos(\Lambda l) - S_{20}\sin(\Lambda l), \quad S_2 = S_{10}\sin(\Lambda l) + S_{20}\cos(\Lambda l), \quad S_3 = S_{30}. \tag{4.10}$$

Equation (3.28) and (4.8) or Eqs. (4.9) and (4.10) describe the polarization variations along the helical ray. The Berry phase (the angle of rotation of the polarization ellipse) per one period of the helix, $l = 2\pi\rho/\sin\vartheta$, equals

$$\Phi_{B0} = -2\pi\cos\vartheta = \Theta - 2\pi, \tag{4.11}$$

where $\Theta$ is the corresponding area on the **t**-sphere, Eq. (3.34) and Fig. 2. An additional $2\pi$ rotation arises from the complete turn of the ray-accompanying coordinate frame after one period of the helix (see Section A1).

## 4.4. Spin Hall effect

Finally, we calculate the polarization-dependent shift of the ray trajectory, $\delta\mathbf{r}$, due to the spin Hall effect. In the Frenet or Euler-angle coordinate frame, the zero-approximation momentum and its derivative, Eqs. (4.1), (4.2), and (4.6), are given by:

$$\mathbf{p}^{(0)} = n\mathbf{t}, \quad \dot{\mathbf{p}}^{(0)} = -n'\mathbf{n}. \tag{4.12}$$

Hence, the polarization term in Eqs. (3.40) is directed along the binormal **b**. Therefore, displacement $\delta\mathbf{r}$ is directed locally orthogonal to inhomogeneity, $\delta\mathbf{p} \equiv 0$, and we can use

Eqs. (3.45) and (3.46) for calculation of $\delta \mathbf{r}$. Substituting Eq. (4.12) into Eq. (3.45) and using condition (4.3), we obtain

$$\delta \dot{\mathbf{r}} = -\lambdabar_0 s \frac{\sin^2 \vartheta}{n\rho} \mathbf{b}, \qquad (4.13)$$

which brings about the shift

$$\delta \mathbf{r} = -\lambdabar s \frac{\sin^2 \vartheta}{\rho} l \mathbf{b}. \qquad (4.14)$$

Note that $\rho^{-1} \sin^2 \vartheta = K^{-1}$ is the curvature of the helical ray, and Eq. (4.14) agrees with Eq. (A12). The shift per one period of helix, $l = 2\pi\rho / \sin\vartheta$, equals

$$\delta \mathbf{r}_0 = -\lambda s \sin\vartheta \mathbf{b}, \qquad (4.15)$$

whereas the splitting between the centers of right-hand, $s = 1$, and left-hand, $s = -1$, polarized beams is, evidently, $\Delta = 2|\delta \mathbf{r}_0|$, see Fig. 4.

## 5. Concluding remarks

We have examined the geometrical optics evolution of light in a gradient-index medium. In the zero-order (scalar) approximation the light propagates along a smooth trajectory which ensures the adiabatic regime of the evolution. The first-order corrections involve the wave polarization and bring about two mutual phenomena – the Berry phase and the spin Hall effect of light. These effects allow different geometrical and dynamical interpretations supplementing one another and equivalent on a deeper level. Here we resume the basic features illuminating the geometrodynamical nature of the Berry phase and spin Hall effect of light.

1. The effects originate from the *spin-orbit interaction* of photons. Polarization and trajectory of light become coupled with each other by the medium inhomogeneity. The Berry phase represents the effect of the trajectory on the polarization of light, while the spin Hall effect is the inverse effect which accounts for the reaction of polarization on the trajectory of propagation. Both the effects are described by a single spin-orbit Lagrangian. The Berry-connection representation unifies the spin-orbit interaction of photons and other quantum particles (e.g., electrons), thereby unveiling their common topological origin.

2. The Berry phase and the spin Hall effect of light can be regarded as manifestations of the effective *Berry vector-potential and field* (geometrically – Berry connection and curvature) minimally coupled to the scalar light. In this manner, the two phenomena are counterparts of the Aharonov–Bohm (Dirac) phase and Lorentz force for a charged particle in a real magnetic field. The spin (helicity) of light plays the role of a charge in the Berry field. Important to note that the Berry curvature appears in *momentum* rather than coordinate space and takes the form of the *topological monopole* as in the case of massless spinning particles.

3. The spin-orbit interaction and the Berry connection arise as a result of the *Coriolis effect* in a non-inertial coordinate frame accompanying the ray trajectory. The Berry phase can be associated with the inertia of the wave field which remains locally non-rotating, and can be eliminated in the proper inertial coordinate frame (see Section A3). In contrast, the spin Hall effect represents a real deflection of the ray trajectory, which is independent on the observation frame.

4. Both the Berry phase and the spin Hall effect are naturally explained in terms of the dynamics of the intrinsic *spin angular momentum* carried by the light. The Coriolis effect is equivalent, in this context, to the *angular Doppler effect* (the medium rotates with respect to the non-inertial frame), whereas the trajectory displacement due to the spin Hall effect is directly related to the *conservation of the total angular momentum* of light. Relations between the Berry phase, spin Hall effect, and the angular momentum of photons are also discussed in [18,19,29].

5. All the above interpretations consider propagation of light as the propagation of a point-like *particle* with some internal properties (spin). However, it is important to remember that we deal with the propagation of a *wave*. The geometrical optics formalism and the notion of the trajectory of propagation imply the evolution of a confined *wave packet* or beam. The Berry phase is the phase of the central plane wave in the packet. At the same time, the spin Hall effect essentially arises from the *interference* of different partial plane waves in the packet, which propagate in slightly different directions and, hence, acquire slightly different geometric phases. In other words, the spin Hall effect appears due to the transverse momentum gradient of the Berry phase [25] (see also [67]). Thus, the spin Hall effect can be attributed to the *uncertainty relation* between the position and momentum – confinement of the wave in the real space inevitably leads to a finite distribution in the momentum space.

Finally, we briefly mention the most important generalizations and extensions of the effects under discussion.

(i) Likewise the angular Doppler shift [72–80] and the Imbert–Fedorov shift [124–129], the Berry phase and spin Hall effect allow immediate generalizations to the case of higher-order beams with vortices bearing intrinsic orbital angular momentum [119,130–135] (see also [84,90]). In this case the intrinsic orbital angular momentum behaves, in the geometrical optics approximation, very much like spin and the vortex charge determines spin-like helicity of the beam. Thus, the spin-orbit-type interaction occurs between the intrinsic and extrinsic orbital angular momentums of light beams, which induces the corresponding Berry phase and orbital Hall effect.

(ii) The geometrodynamical description of the propagation of light in gradient-index media can be extended to the case of weakly anisotropic media [34,35,136–138]. In this manner, the effect of anisotropy is described by the generalized $SO(3)$ evolution of the Stokes vector on the Poincaré sphere with the non-conserved helicity.

(iii) The Berry phase and spin Hall effect for smooth trajectories of propagation have a universal character, and appear not only for light in a gradient-index medium, but also for evolution of relativistic particles in external potentials [17–19,29,32,33,54], propagation of transverse elastic waves in inhomogeneous media [139], propagation of light in gravitational fields [30], photonic crystals [27,28], and metamaterials [140].

(iv) A number of effects related to the spin-orbit coupling of light take place in multimode optical fibers [24,120–123,141–144].

(v) Finally, the spin-orbit interaction of light and spin-to-orbit conversion of the angular momentum (which can be associated with the spin Hall effect) occurs upon the scattering of polarized plane waves on various non-planar objects, such as lenses or small scatterers, or in inhomogeneous anisotropic media [67,145–152]. Anisotropy per se may also induce spin-orbit coupling upon the propagation of confined beams even in a homogeneous medium [153,154].

The tiny optical phenomena induced by the spin-orbit interaction of light are anticipated to have promising applications in nano-photonics. The spin Hall effect with a typical scale of the order of the wavelength provides a novel type of transport which can be crucial for modern optics dealing with nano-scales. The dynamics and geometry are strikingly entangled in the spin-orbit coupling phenomena, which unveils the fundamental intrinsic properties of photons. The complementarity between the Berry phase and spin Hall effect (as well as the complementarity between the Aharonov-Bohm phase and the Lorentz force) is closely related to the duality of the wave and particle aspects of semiclassical evolution.

# Acknowledgments

I am indebted to M.V. Berry, P.A. Horváthy, C. Duval, Y.A. Kravtsov, C.N. Alexeyev, and A.V. Volyar for fruitful discussions. This work was partially supported by the Australian Research Council through the Linkage International grant.

# Appendix A: Ray coordinate frames

Let the laboratory coordinate frame be $(XYZ)$ with basis vectors $(\mathbf{e}_X, \mathbf{e}_Y, \mathbf{e}_Z)$. To describe the polarization of a transverse wave, one needs to use a co-moving coordinate frame $(xyz)$ accompanying the scalar-approximation ray trajectory (2.5). The $z$ axis of such frame is directed along the wave momentum, so that the basis vectors are $(\mathbf{v}, \mathbf{w}, \mathbf{t})$. There is a freedom of choice of directions $(\mathbf{v}, \mathbf{w})$, which are determined up to an $SO(2)$ rotation in the $(xy)$ plane, and here we address several specific ray-accompanying frames which can be useful for different applications.

## A.1. Euler-angles frame

To superimpose the z-axis and a given direction of the local wave momentum, $\mathbf{t}$, one needs to perform a local rotation of the coordinate frame. The simplest way to do this is to use the Euler angles and rotational operators describing $SO(3)$ rotations in space. Let the direction of $\mathbf{t}$ be given by spherical angles $(\theta, \phi)$, Fig. 2, i.e., in the laboratory coordinate frame $\mathbf{t} = (\sin\theta\cos\phi, \sin\theta\sin\phi, \cos\theta)$. The desired rotation transformation can be performed as follows:

$$\begin{pmatrix} \mathbf{v} \\ \mathbf{w} \\ \mathbf{t} \end{pmatrix} = \hat{R} \begin{pmatrix} \mathbf{e}_X \\ \mathbf{e}_Y \\ \mathbf{e}_Z \end{pmatrix}, \tag{A1}$$

where the rotational matrix $\hat{R}$ consists of two successive rotations and equals:

$$\hat{R} = \hat{R}_X(\theta) \hat{R}_Z\left(\frac{\pi}{2} + \phi\right) = \begin{pmatrix} -\sin\phi & \cos\phi & 0 \\ -\cos\theta\cos\phi & -\cos\theta\sin\phi & \sin\theta \\ \sin\theta\cos\phi & \sin\theta\sin\phi & \cos\theta \end{pmatrix}. \tag{A2}$$

Here $\hat{R}_A(\gamma)$ is the operator of rotation on angle $\gamma$ about the $A$ axis. Matrix $\hat{R}$ corresponds to a rotation determined by the Euler angles $(\phi_E, \theta_E, \psi_E) = \left(\frac{\pi}{2} + \phi, \theta, 0\right)$ [155]. Therefore, we refer to this $(\mathbf{v}, \mathbf{w}, \mathbf{t})$ coordinate frame as to the *Euler-angles frame*. Transformation similar to Eqs. (A1) and (A2) has been used in [31,34] for diagonalization of Maxwell equations (3.1) and in [79] for explanation of the angular Doppler shift.

From Eqs. (A1) and (A2) we have:

$$\begin{aligned} \mathbf{v} &= -\mathbf{e}_X \sin\phi + \mathbf{e}_Y \cos\phi, \\ \mathbf{w} &= -\mathbf{e}_X \cos\theta\cos\phi - \mathbf{e}_Y \cos\theta\sin\phi + \mathbf{e}_Z \sin\theta, \\ \mathbf{t} &= \mathbf{e}_X \sin\theta\cos\phi + \mathbf{e}_Y \sin\theta\sin\phi + \mathbf{e}_Z \cos\theta. \end{aligned} \tag{A3}$$

Substituting Eq. (A3) to Eq. (3.5), we obtain the angular velocity of the Euler-angles frame with respect to the laboratory one [155]:

$$\boldsymbol{\Omega} = \mathbf{t}\dot{\phi}\cos\theta + \mathbf{v}\dot{\theta} + \mathbf{w}\dot{\phi}\sin\theta, \quad \Omega_\parallel = \dot{\phi}\cos\theta. \tag{A4}$$

Equation (3.25) yields the expression for the Berry phase in the Euler-angles frame [16,31,34]:

$$\Phi_B = -\int \cos\theta \, d\phi. \tag{A5}$$

For a closed loop on the **t**-sphere, this phase coincides with the Berry phase (3.33) and (3.34) up to $2\pi$ difference. This difference is caused by the rotation of the Euler-angles frame. Indeed, let $\theta \to 0$ and $\phi$ varies from 0 to $2\pi$, which corresponds to a nearly rectilinear wave propagation along the $Z$ axis and a nearly zero-area contour on the **t**-sphere, Fig. 2. It can be seen, that in this case the Euler-angles frame performs a complete $2\pi$ turn in the $(XY)$ plane.

The Berry connection and curvature can be calculated by substituting Eqs. (A3) into Eqs. (3.17)–(3.18') and differentiating basis vectors in spherical coordinates $(p,\theta,\phi)$ in the **p** space. As a result, we obtain

$$\mathbf{A} \equiv (A_p, A_\theta, A_\phi) = p^{-1}(0, 0, -\tan^{-1}\theta), \quad \mathbf{F} \equiv (F_p, F_\theta, F_\phi) = p^{-2}(1, 0, 0). \tag{A6}$$

Here the Berry curvature is the monopole (3.20), whereas the Berry connection becomes Eq. (3.35) after gauge transformation (3.23) with $\alpha = -\phi$.

The Euler-angles frame is convenient for explicit analytic calculations. At the same time, its weakness is that it depends on the choice of the laboratory frame $(XYZ)$.

*A.2. Frenet frame*

An alternative coordinate frame, accompanying geometrical optics rays is the *Frenet frame* widely used in differential geometry of curves [156]. It is based on the principal normal **n**, binormal **b**, and tangent **t** to the ray:

$$(\mathbf{v}, \mathbf{w}, \mathbf{t}) = (\mathbf{n}, \mathbf{b}, \mathbf{t}). \tag{A7}$$

These basis vectors characterize local properties of the curve and are independent on the laboratory coordinate frame. They can be defined as

$$\mathbf{t} = \frac{\dot{\mathbf{r}}}{|\dot{\mathbf{r}}|}, \quad \mathbf{n} = \frac{\dot{\mathbf{t}}}{|\dot{\mathbf{t}}|}, \quad \mathbf{b} = \mathbf{t} \times \mathbf{n}. \tag{A8}$$

Note that vectors (A8) depend not only on the direction **t**, but also on its derivative $\dot{\mathbf{t}}$.

The Frenet-Serret equations describe evolution of the frame along the ray [59,156]:

$$\dot{\mathbf{t}} = K^{-1}\mathbf{n}, \quad \dot{\mathbf{n}} = -K^{-1}\mathbf{t} + T^{-1}\mathbf{b}, \quad \dot{\mathbf{b}} = -T^{-1}\mathbf{n}. \tag{A9}$$

Here $K^{-1}$ and $T^{-1}$ are, respectively, the curvature and torsion of the ray. Comparing these equations with Eq. (3.4) we find the local angular velocity of rotation of the Frenet frame:

$$\mathbf{\Omega} = T^{-1}\mathbf{t} + K^{-1}\mathbf{b}, \quad \Omega_\parallel = T^{-1}. \tag{A10}$$

Thus, the torsion characterizes local rotation of this frame about **t**, and the Berry phase in this frame is

$$\Phi_B = -\int T^{-1} dl. \tag{A11}$$

The geometrical phase in this form is equivalent to the Rytov's law of the polarization evolution, which had been invented long before the Berry phase discovery [4,8,9,59].

The Berry connection cannot be introduced in the Frenet frame in the generic case, because of the local dependence of its basis vectors (A8) on derivative $\dot{\mathbf{t}}$. But for closed contours the integral (A11) coincides with the phase (3.33) and (3.34) up to a $2\pi$ difference.

The Frenet frame is particularly convenient for description of the spin Hall effect. Indeed, the topological term in equations (3.40) can be written as [35]

$$\bar\lambda_0 s \frac{\dot{\mathbf{p}} \times \mathbf{p}}{p^3} = \bar\lambda s \dot{\mathbf{t}} \times \mathbf{t} = -\bar\lambda s K^{-1}\mathbf{b}. \tag{A12}$$

This shows that the spin-induced ray deflection is determined solely by the ray curvature and is directed along the binormal, i.e., orthogonal to the local propagation plane $(\mathbf{t}, \mathbf{n})$. Thus, while the Berry phase is an artifact of a particular rotating ray-accompanying coordinate frame, the spin-Hall effect is a Coriolis-type transverse deflection induced by a real bending of spinning light.

## *A.3. Parallel-transport frame*

Among all the possible ray-accompanying frames, there is a special frame which should be distinguished from others. This is a non-rotating (with respect to $\mathbf{t}$), i.e., locally *inertial* frame with

$$\Omega_\parallel = 0. \tag{A13}$$

In other words, the basis vectors of this frame, $(\mathbf{u}_1, \mathbf{u}_2, \mathbf{t})$, are parallel transported along the ray, and therefore it is called the *parallel-transport frame*. Its basis vectors can be introduced by a rotation of another ray-accompanying coordinate frame on angle $\Phi_B$. For instance, starting from the Frenet frame $(\mathbf{n}, \mathbf{b}, \mathbf{t})$, basis vectors $(\mathbf{u}_1, \mathbf{u}_2, \mathbf{t})$ can be introduced as

$$\begin{pmatrix} \mathbf{u}_1 \\ \mathbf{u}_2 \end{pmatrix} = \begin{pmatrix} \cos\Phi_B & \sin\Phi_B \\ -\sin\Phi_B & \cos\Phi_B \end{pmatrix} \begin{pmatrix} \mathbf{n} \\ \mathbf{b} \end{pmatrix}, \tag{A14}$$

where $\Phi_B = -\int T^{-1} dl$. Substituting Eq. (A14) into Frenet–Serret equations (A9), we arrive at

$$\dot{\mathbf{t}} = \frac{\mathbf{u}_1}{K}\cos\Phi_B - \frac{\mathbf{u}_2}{K}\sin\Phi_B, \quad \dot{\mathbf{u}}_1 = -\frac{\mathbf{t}}{K}\cos\Phi_B, \quad \dot{\mathbf{u}}_2 = \frac{\mathbf{t}}{K}\sin\Phi_B. \tag{A15}$$

The last two equations (A15) are equivalent to the parallel-transport equation (3.32) [156]:

$$\dot{\mathbf{u}}_1 = -(\mathbf{u}_1 \dot{\mathbf{t}})\mathbf{t}, \quad \dot{\mathbf{u}}_2 = -(\mathbf{u}_2 \dot{\mathbf{t}})\mathbf{t}, \text{ so that } \mathbf{u}_i \dot{\mathbf{u}}_j = 0. \tag{A16}$$

Using Eq. (3.5), the angular velocity of the parallel-transport frame equals

$$\boldsymbol{\Omega} = -\frac{\sin\Phi_B}{K}\mathbf{u}_1 + \frac{\cos\Phi_B}{K}\mathbf{u}_2, \quad \Omega_\parallel = 0. \tag{A17}$$

Thus, although the rotation of the ray-accompanying frame about $\mathbf{t}$ can be eliminated, rotation associated with the ray bending and the ray curvature is unremovable. It is this rotation that causes the spin Hall effect [135].

The Berry phase vanish in this frame. However, the parallel-transport frame is *non-local* – equation (A14) contains the phase which depends on the integral evolution of $\mathbf{t}$. Therefore, this system cannot be parametrized by the local value of $\mathbf{t}$. Also, a transition from a local ray-accompanying frame to the parallel-transport frame *cannot* be considered as a local gauge transformation (3.21)–(3.24) which keeps the Berry curvature and phase invariant. It is worth noticing that a curvilinear coordinate systems based on unit vectors $(\mathbf{u}_1, \mathbf{u}_2, \mathbf{t})$ is the only *orthogonal* coordinate system in the vicinity of the ray trajectory [81,87,157].